\documentclass[runningheads]{llncs}
\usepackage{graphicx}
\usepackage{placeins}
\usepackage{siunitx}
\usepackage{comment}

\usepackage{amsmath}
\usepackage[inline]{enumitem}
\usepackage{booktabs}
\usepackage{makecell}
\usepackage{float}
\usepackage{subcaption}
\usepackage{multirow} 
\usepackage[hyphens]{url}
\usepackage{array}
\newcommand{\PreserveBackslash}[1]{\let\temp=\\#1\let\\=\temp}
\newcolumntype{C}[1]{>{\PreserveBackslash\centering}p{#1}}
\newcolumntype{R}[1]{>{\PreserveBackslash\raggedleft}p{#1}}
\newcolumntype{L}[1]{>{\PreserveBackslash\raggedright}p{#1}}

\begin{document}

\title{
The Cyber Alliance Game: How Alliances Influence Cyber-Warfare}


\author{Gergely Benkő\inst{1} \and Gergely Biczók\inst{1,2,3}}

\authorrunning{G. Benkő, G. Biczók}

\institute{CrySyS Lab, Dept. of Networked Systems and Services,\\ Budapest University of Technology and Economics \\
\email{benkogergely@edu.bme.hu, biczok@crysys.hu}
\and
BME-HUN-REN Information Systems Research Group, Hungary
\and
University of Michigan, Ann Arbor, MI 48109, USA
}

\maketitle          
\setcounter {footnote}{0}
\begin{abstract}
Cyber-warfare has become the norm in current ongoing military conflicts. Over the past decade, numerous examples have shown the extent to which nation-states become vulnerable if they do not focus on building their cyber capacities. Adding to the inherent complexity of cyberwar scenarios, a state is usually a member of one or more alliances. Alliance policies and internal struggles could shape the individual actions of member states; intuitively, this also holds for the cyber domain.

In this paper, we define and study a simple Cyber Alliance Game with the objective of understanding the fundamental influence of alliances on cyber conflicts between nation-states. Specifically, we focus on the decision of whether to exploit a newly found vulnerability individually or share it with the alliance. First, we characterize the impact of vulnerability-sharing rewards on the resulting equilibrium. Second, we study the implications of the internal power structure of alliances on cyberwar outcomes and infer the expected behavior of Dictator, Veto, and Dummy players. Finally, we investigate how alliances can nudge their members via rewards and punishments to adhere to their defensive or offensive cyber policy. We believe that our results contribute to the fundamental understanding of real-world cyber-conflicts by characterizing the impact of alliances.

\keywords{Cyber-Warfare \and Weighted Voting \and Game Theory \and Alliances}
\end{abstract}

\section{Introduction}
\label{sec:introducion}
The ongoing Russo-Ukrainian war highlights the importance of military cyber operations and the significant difference between cyber- and traditional warfare in its dynamics, points of conflict, and applied weaponry~\cite{robinson2015cyber}. Cyberattacks can be broadly categorized into two types: they could target either military or civilian objects, e.g., weapon guidance systems\footnote{\url{https://www.wired.com/story/dire-possibility-cyberattacks-weapons-systems/}} or telecommunication companies such as Kyivstar~\cite{santora2023huge} (see Appendix~\ref{sec:real} for examples from both sides). 

While the decision situations in an ongoing cyber conflict can be complex in themselves, most countries are also part of military and/or economic alliances, which have strict admission criteria, internal rules, and a common purpose. Once a nation-state becomes a member of an alliance, it represents both its own interests and those of the alliance through its actions and behavior. During the collaboration, members align their military and economic objectives; nowadays, a common cyber defense strategy has also been implemented in most alliances, blurring the lines between military and other alliances. The European Union has been advocating for harsh sanctions against cybercrime and supporting joint research and knowledge sharing \cite{christou2018challenges}. BRICS+ countries research three main topics within the framework of the CyberBRICS project~\cite{belli2021cyberbrics}: protection of personal data, secure digital transition, and regulation of artificial intelligence~\cite{cyberbrics}. While NATO has prioritized the defense of information systems ~\cite{hunker2010cyber} and employs a policy of deterrence through cyber exercises and capability enhancement, the operations, regulations, and guidelines of the Collective Security Treaty Organization (CSTO), which includes post-Soviet successor states, receives less public attention regarding cyberspace~\cite{elamiryan2018comparing}. (See Appendix~\ref{sec:alliances} for more details.)

Military operations have been a fertile application domain for studying strategic decision-making via game theory since its inception~\cite{BeckmannReimer+2014+193+215}. Recently, cyber-warfare has also been studied with a similar toolset, but existing studies have not considered alliance systems and their impact on the actors. For instance, in their seminal work~\cite{moore2010would}, the authors modeled and studied the discovery process, stockpiling, and weaponization of (software) vulnerabilities in the cyber-conflict of two nation-states. Their main finding was that at least one state has the incentive to act as an aggressor as opposed to focusing on defensive efforts in a wide range of scenarios.

In this paper, we ask the question: \emph{how do the presence of alliances influence the outcomes of cyber-warfare?}. Inspired by the Cyber Hawk game~\cite{moore2010would}, we construct and analyze the equilibria of the \emph{Cyber Alliance Game (CAG)}. Specifically, through games of increasing complexity and with the help of weighted voting theory~\cite{nordmann1999weighted}, we show how i) alliances can encourage peaceful, defensive outcomes via rewards for sharing newly discovered vulnerabilities, ii) the internal power structure within an alliance affects the equilibria, and iii) how the fundamental posture of alliances (defensive or offensive) perturb the resulting strategy profiles. Although our model is quite generic and high-level, we believe that our results can help better understand and perhaps predict real-world cyber-warfare-related decisions of various member states of different alliances.

The rest of the paper is structured as follows. Section \ref{sec:background} lays out the terminology, the basics of weighted voting, and related modeling efforts. Section \ref{sec:CAG} defines our baseline game CAG and analyzes its possible outcomes. Section \ref{sec:power} extends CAG with intra-alliance power structures. Section \ref{sec:policy} further extends the model with the fundamental defensive/offensive posture of alliances. Finally, Section \ref{sec:conclusion} provides a discussion on the implications of our results and concludes the paper.

\section{Background}
\label{sec:background}
\subsection{Terminology}
Here, we define the basic cybersecurity concepts used throughout the paper. A \textit{vulnerability} constitutes an opportunity for attack. If previously undiscovered or undisclosed, it can be referred to as a \emph{zero-day}. Software vulnerabilities are the most common, but there also exist hardware, organizational, and human vulnerabilities. An \textit{exploit} is a weaponized vulnerability; the attacker develops a piece of code with which he can corrupt the system through the identified vulnerability. A discoverer of a specific vulnerability can never be certain whether they are the sole possessor of that piece of knowledge (unless it was made public or they were attacked through that exact vulnerability), it is possible that they just made a \textit{rediscovery}~\cite{ozment2006milk}.

Nevertheless, a strategic discoverer can choose between several actions~\cite{moore2010would}. First, they can \textit{stockpile}, storing a discovered vulnerability for later offensive usage.
Note that alleged zero-days can also be acquired from the black market, resulting in a similar situation~\cite{allodi2017economic}.
Second, they can \textit{patch} their own system via fixing the vulnerability, but at the cost of eliminating the possibility of an attack (patching makes the vulnerability public, hence other actors can also patch their system, resulting in an overall increased level of cybersecurity). The actual patch development is usually done by software vendors~\cite{rescorla2005finding}, but a nation-state might also be able to fix critical vulnerabilities. A glimpse into the economics of a real-world vulnerability disclosure process and related actions of the United States was given in~\cite{caulfield2017us} based on a redacted version of the document and a blog post by the former White House Cybersecurity Coordinator. Finally, in relation to alliances, an actor can choose to \textit{share} the vulnerability results with its alliance. With this step, the discoverer forfeits the opportunity for individual exploitation and places the decision in the hands of the alliance and its internal processes.

\subsection{Game theoretical models of cyber-warfare}
\label{sec:hawk}

Although there has been much research in modeling different aspects of attacker-defender dynamics in the cyber domain~\cite{cybergt_survey,HUNT2024401}, the cyber struggle specific to nation-states has received considerably less attention in the game theory literature. In the seminal paper already mentioned above~\cite{moore2010would}, Moore et al. study two simple cyberwar games (one on stockpiling and one on weaponizing/exploiting the discovered vulnerability) between two nation-states. The factors they take into account are the technical sophistication of the states and their willingness to attack. Although the equilibrium outcomes are influenced by these factors, the key takeaway is that there is no all-defensive equilibrium: at least one of the players (whether based on perceived technical superiority or a more aggressive posture) will choose stockpile/attack.

Next, Bao et al. present a comprehensive game-theoretical framework to be used as an automated tool in cyber-strategy development~\cite{bao_cyber}. This paper relaxes the many simplifying assumptions of previous works and develops a computationally feasible algorithm for deriving equilibria in the complex model. Focusing on algorithmic details, this paper does not offer high-level insights on cyber-warfare.

Chen et al.\cite{chen2020disclose} introduced a practical framework for governments that, utilizing various parameters, rapidly determines whether the attacker's exploitation of a vulnerability or patching it would yield greater benefits. If a country's systems are affected or if the manufacturer is located within the territory of that country, patching becomes significantly more cost-effective. Another important finding (in line with~\cite{moore2010would}) is that technically inferior countries are more likely to opt for publicly disclosing the vulnerability. Finally, Wang et al. focused on a two-player two-stage conflict~\cite{wang_hausken}, where one player can both invest into attacking and stockpiling in the first stage, while it can also use the stockpile created in the second stage; the other player defends in both stages. The paper studied how players managed the trade-off in exerting their efforts across the two periods depending on asset valuations, asset growth, time discounting, and contest intensities.

As opposed to the related literature, we analyze simple two-player cyber-warfare games inspired by ~\cite{moore2010would} and focus exclusively on the impact of alliances. This approach enables us to i) compare to a baseline without alliances and ii) isolate the effect of alliances, adding a new dimension to the strategic understanding of cyber-warfare situations.


\subsection{Weighted voting}
Weighted voting is a crucial mechanism for making collective decisions~\cite{nordmann1999weighted}, especially in environments where various stakeholders have differing levels of expertise, resources, and risk exposure. Weighted voting assigns different weights to the votes of different participants, reflecting their relative importance, expertise, or investment in the outcome~\cite{weighted}. In the context of (cyber) alliances, such mechanisms are widely used to establish the joint decisions of the member states.

\noindent \textbf{Terminology. }Each entity casting a vote is called a \textit{Player} in the election. They’re often denoted as $P_1$,$P_2$,$P_3$,...,$P_N$  where $N$ is the total number of voters. A player is given a \textit{Weight}, which represents how many votes they get. The \textit{Quota} is the minimum aggregated weight needed for the votes to be approved; the quota must be larger than $50\%$ and not larger than $100\%$ of the total number of votes. The \textit{Coalition} is a group of players who voted in the same way. A coalition is a \textit{Winning coalition} if it has enough weight to reach the quota. A Player is \textit{Critical} if the coalition loses the vote without them. A weighted voting system is often represented in a shorthand form: [$q$:$w_1$,$w_2$,$w_3$,...,$w_N$]. In this form, $q$ is the quota, while $w_i$ is the weight of Player $i$.

\noindent \textbf{Power relations. }Players of different powers may fall into distinct categories of importance. A player is a \textit{Dictator} if their weight is equal to or greater than the quota. A player has \textit{Veto} power if their support is essential to reach the quota. Finally, a player is a \textit{Dummy} if they are never essential to reach the quota. In general, the power of a player can be characterized by the Banzhaf Power Index that determines the proportion of all winning coalitions where the given player is critical, analogously to the Shapley-Shubik index in cooperative games~\cite{strafiin1988shapley}.

\section{The Cyber Alliance Game}
\label{sec:CAG}
 Cyber conflicts entail the risk of escalation, but they also present a completely new opportunity for the aggressor. If there is reason to believe that escalation is unlikely, cyber conflicts offer an opportunity to increase the influence of nation-states. During the strategic decision-making process, the cyber commander must leverage their belief in the likelihood of the adversary exploiting a specific vulnerability.  The statutes of most alliances condemn private actions and may even introduce sanctions against members, tarnishing the reputation of the alliance. Nevertheless, the members of these alliances are diverse, autonomous entities driven by entirely different objectives and motivations.
 
 Below, we define the Cyber Alliance Game (CAG); our investigation revolves around how the alliance environment influences the decisions of two opposing nation-states belonging to different alliances. The model deals with the costs and benefits of exploiting or sharing vulnerabilities without considering the outcomes of an actual conflict; as such, CAG is inspired by the Cyber Hawk game and adopts its base parameters ($p$ and $q$, see later) and mechanics regarding how the game is played~\cite{moore2010would}. The game is played by two players who are members of two different alliances.

\subsection{Strategies and game mechanics}

Each player can choose from two strategies.

\noindent \textbf{Attack. } The player discovers a vulnerability and develops an exploit from it but keeps it secret from the alliance (and the world). An attack is considered successful when a player discovers a vulnerability and uses it before any other player does. The reward for being the first to launch a successful attack is 1, while being subjected to an attack results in a cost of -1, making this a zero-sum strategy.

\noindent \textbf{Share. }The player discovers a vulnerability and shares it with their allies. The alliance then decides whether to develop an exploit and collectively utilize it or to forward it to the vendor~\cite{chen2020disclose}. To preserve clarity, decision-making after sharing the vulnerability is omitted from CAG.
The first player who shares a specific vulnerability with their alliance receives a payout; the alliance's reward to the Player. In this situation, the other player receives a payout of $0$, making sharing a non-zero-sum strategy.

\begin{figure}[!b]
	\centering
	\includegraphics[width=0.63\textwidth]{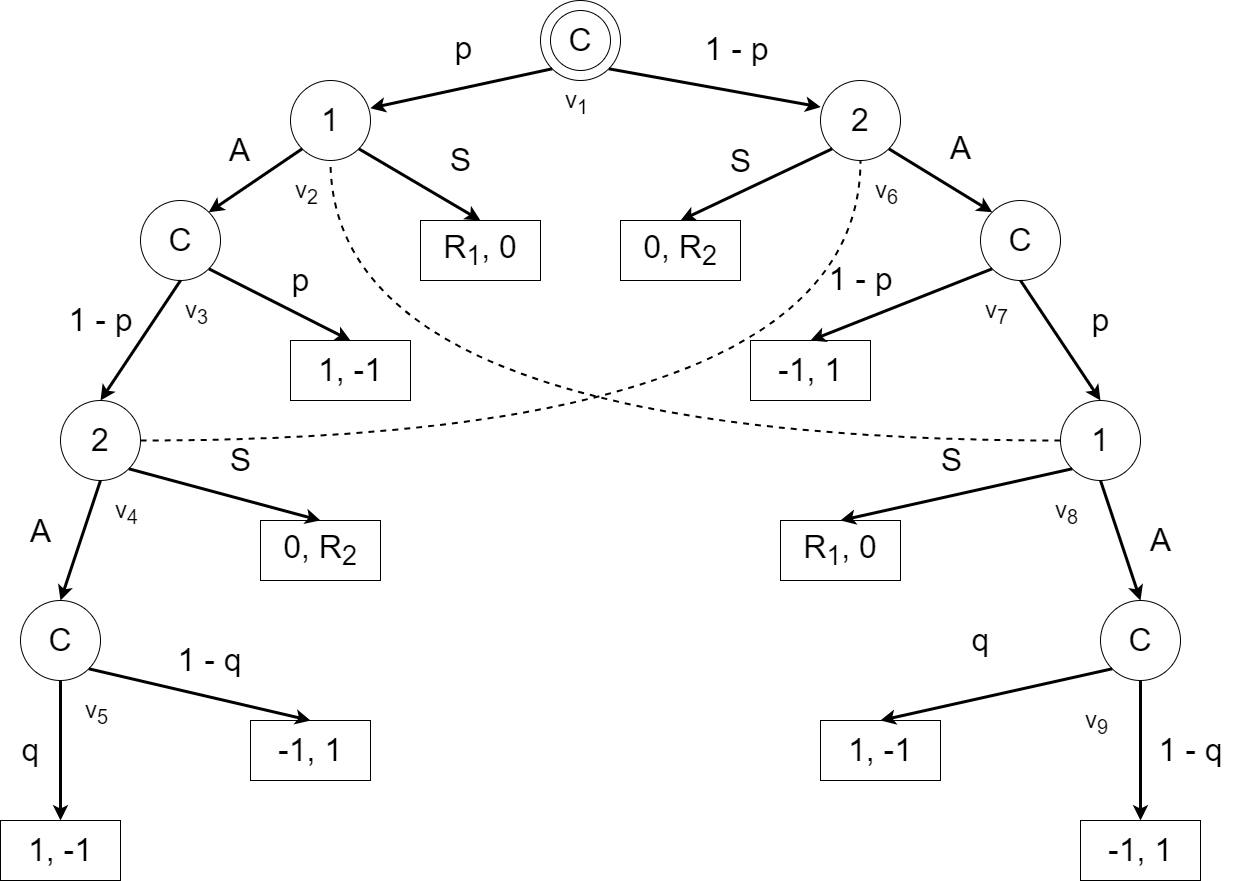}
	\caption{Cyber Alliance Game in extensive form}
	\label{fig:game1}
\end{figure}

Using Fig.~\ref{fig:game1}, we illustrate the mechanics of the game. Decision nodes are represented by circles, and leaf nodes by rectangles. Nodes can be labeled with either the letter ``C'' for chance or the number of the player. A numbered node and its children have two edges labeled ``A'' for Attack and ``S'' for Share, denoting the two possible moves. The edges between chance nodes and their children are labeled with probabilities. At the leaves of the tree, the first number in the rectangle represents the payoff for Player 1, while the second number represents the payoff for Player 2. The game starts at the root of the tree, indicated by a double circle.

\subsection{Game parameters}

Three parameters influence the payoffs and, thus, the equilibrium of CAG. These are the following:

\noindent \textbf{p.} The parameter $p$ takes a value between $0$ and $1$ and measures the player's technical savvy in discovering vulnerabilities. The value of $1-p$ represents the technical sophistication of the other player. Smaller values of $p$ indicate that Player 1 is less sophisticated than Player 2, while larger values indicate the opposite. If the two players are evenly matched, then $p = 0.5$. CAG interprets $p$ as the probability of discovering a new vulnerability. 

\noindent \textbf{q. }The parameter $q$ measures the willingness of each player to launch an attack after discovering a vulnerability. Its value ranges from $0$ to $1$ and indicates how aggressive a player is. The value of $1-q$ represents the attack willingness of the other player. Smaller values of $q$ indicate that Player 1 is more restrained in launching attacks, while larger values indicate that Player 1 is ``trigger-happy''. If the two players are evenly matched, then $q = 0.5$.

\noindent \textbf{R$_i$. }The $R_i$ value represents the alliance's reward to a member who shared a valuable vulnerability. The value of $R_i$ also falls between $0$ and $1$. In real-world scenarios, various factors can affect the magnitude of alliance rewards, such as the exploitability of the vulnerability or the number of affected systems. By introducing this reward, CAG becomes a non-zero-sum game (unlike the Cyber Hawk game) and enables modeling the influence of the alliance environment on the player's decision regarding the discovered vulnerability. 

\subsection{Payoffs}

To find a Nash equilibrium, we need to determine the payoff for each player for every possible strategy profile. A strategy profile is referred to as the pair ($S_1$, $S_2$), where S$_1$ represents the first player's strategy, and S$_2$ represents the second player's strategy. The corresponding expected payoffs are denoted as $u_i$(S$_1$, S$_2$).

The expected payoffs are determined following the game mechanics. 
At nodes labeled with ``C'', the probability of reaching a specific child is associated with the value on the corresponding edge. At numbered nodes, the respective player must choose between the actions Attack (A) and Share (S). After making the choice, the game proceeds along the edge labeled with the chosen action. Leaf nodes signify the end of the game: the players receive the payoff specified in the leaf.

Let's determine the payoff for Player $1$ when it chooses Attack (A), and their opponent chooses Share (S): $u_1$(A, S).
The game starts at the root, which is the node $v_1$. Initially, the game proceeds along the left branch with a probability of p. Upon reaching the node $v_2$, Player $1$ is faced with a decision and chooses to Attack (A). After this choice, we arrive at the chance node $v_3$, where the game ends with a probability of p. In this case, the first Player receives a payoff of 1. Otherwise, With a probability of $1-p$, Player $2$ rediscovers the vulnerability, leading to node $v_4$. At node $v_4$, the Player $2$ chooses the Share (S) action, resulting in a payoff 0 for the first Player. Thus we get:
\begin{align}u_1(A,S)=p \cdot (p*1+(1-p) \cdot 0) + (1-p) \cdot 0= p \cdot (p \cdot 1) =p^2\nonumber\end{align}

Note that at the root node, Player 2 discovers the vulnerability first with probability $1-p$; in this case, the game progresses along the right branch instead.

It is important to emphasize that player nodes belong to two distinct information sets ($v_2$ and $v_8$ versus $v_4$ and $v_6$, see the dashed lines in Fig.~\ref{fig:game1}). Owing to imperfect information, a player does not know which exact node they are at inside their own information set; they do not know whether they are the first to discover a specific vulnerability.

Following the method above, we determine the expected payoff for each strategy profile:
\begin{align}
u_1(A,S)&= p^2\nonumber\\
u_1(S,A)&=-p2 +2 \cdot p \cdot R_1 +2p - p^2 R_1  -1\nonumber\\
u_1(S,S) &=p \cdot R_1 \nonumber\\
u_1(A,A)&=2p^2 + 4pq -4p^2q -1\nonumber\\
u_2(A,S)&=-p2 -p2R_2 +R_2\nonumber\\
u_2(S,A)&=(1-p)^2 \nonumber\\
u_2(S,S)&=(1-p)*R_2\nonumber\\
u_2(A,A)&=-2p^2-4pq+4p^2q+1\nonumber.
\end{align}
Note that in the case of the $(A,A)$ strategy profile, the expected payoff depends on who attacks first (Player 1 with a probability of $q$ and Player 2 with a probability of $1-q$).

Afterward, we can represent the game in a 2-by-2 matrix format (see Table \ref{tab:matrix}). In the example, we use the following parameter values: $p=0.6$, $q=0.3$, $R1=0.7$, $R2=0.45$.

\begin{table}[tb]
    \centering
    \begin{tabular}{|c|c|c|} \hline 
         P$_1$/P$_2$&  S& A\\ \hline 
         S&  0.42,0.18& 0.428,0.16\\ \hline 
         A&  0.36,-0.072& 0.008,-0.008\\ \hline
    \end{tabular}
    \vspace{1mm}
    \caption{CAG payoffs, $p=0.6$, $q=0.3$, $R1=0.7$, $R2=0.45$.}
    \label{tab:matrix}
\end{table}

\subsection{Equilibrium analysis}

Similarly to \cite{moore2010would}, we utilize simple (pure strategy) Nash Equilibrium (NE) as our solution concept (as opposed to more involved sequential equilibrium concepts~\cite{sequential}). We can do this because a strategy in an extensive-form game with imperfect information is a function that maps information sets to actions; in fact, all the nodes in an information set have identical available actions. In CAG, this means that a player's strategy is either A or S, and the same action is played in either node of a given player.

In CAG, there are three parameters, each of which can take a value between $0$ and $1$. For solving and visualizing CAG instances, we wrote a Python script\footnote{Available at \url{https://anonymous.4open.science/r/cyberwar-8F21/}}. Specifically, we generate NE diagrams where the x-axis represents $p$, the y-axis represents $q$, and the $R_i$ values are fixed according to various criteria.


\begin{figure}[tb]
	\centering
	\includegraphics[width=40mm, keepaspectratio]{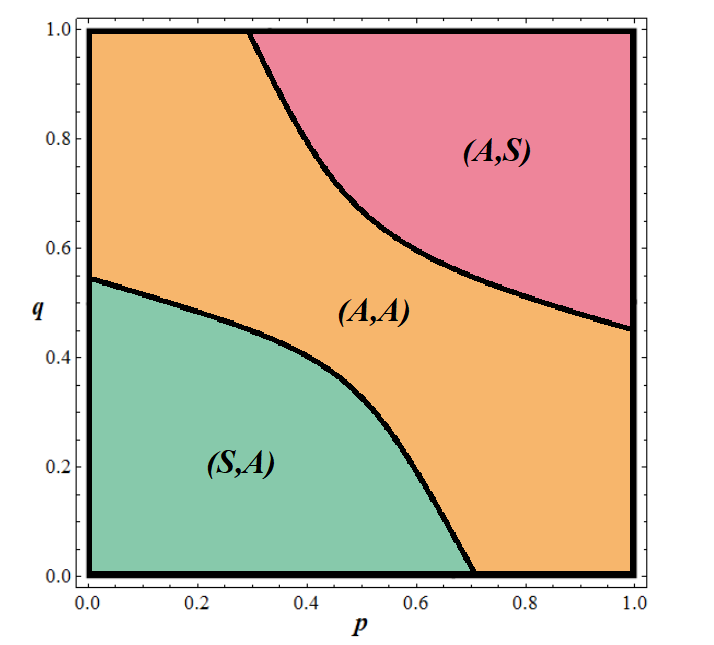}
	\includegraphics[width=40mm, keepaspectratio]{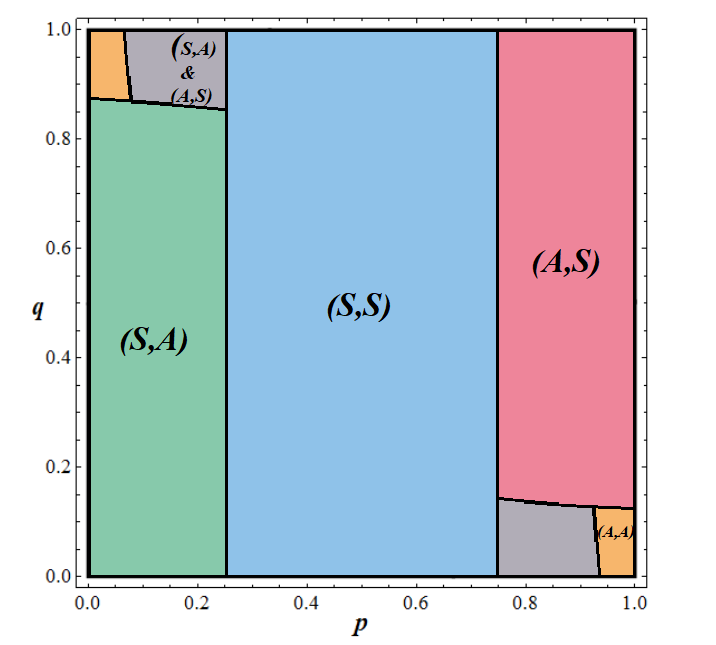}
 \includegraphics[width=40mm, keepaspectratio]{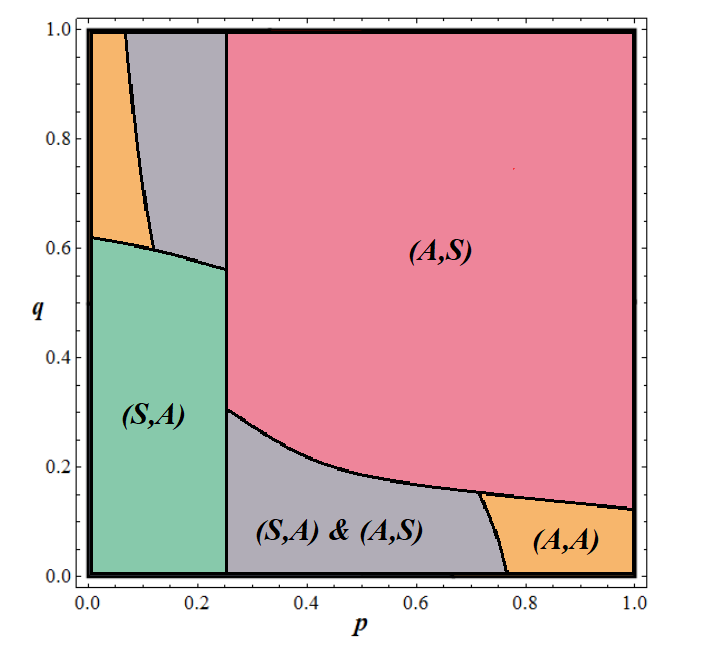}\\
 \vspace{-3mm}
 \caption{Left: $R_1$=0.1, $R_2$=0.1; Mid: $R_1$=0.75, $R_2$=0.75; Right: $R_1$=0.25, $R_2$=0.75 }
	\label{fig:cag1}
 \vspace{-5mm}
\end{figure}

If both players receive equally low rewards from their alliance for sharing the discovered vulnerability, they tend to choose attack for a wide range of ($p$,$q$) values (left plot in Fig.~\ref{fig:cag1}). As rewards go towards zero, CAG essentially transforms into the Cyber Hawk game~\cite{moore2010would}. Moreover, if the players' technical proficiency (p) and aggressiveness (q) are similar, they both opt for attack (orange), resulting in a risk of conflict escalation. Since most real-world alliances are defensive, it is in their best interest to offer high(er) rewards for vulnerability sharing to maintain peace.

As we begin to concurrently increase the rewards above a value of 0.5, the (S,S) equilibrium emerges (blue), promoting sharing in case of similar technical proficiency $p$ (middle plot in Fig.~\ref{fig:cag1}). As we approach maximum symmetric rewards, states will share vulnerabilities for any ($p$,$q$), trusting the alliance with the ultimate decision concerning a cyberattack.

If alliance rewards are highly asymmetric, the player with the higher reward is more likely to play S, while the under-rewarded player will attack (red (A,S) NE, the right plot in Fig.~\ref{fig:cag1}).
Note that the behavior of an alliance member can be predicted from the alliance's rewards. This enables the other player (if familiar with this information) to optimize its own payoff; treating rewards as top-secret is advised.


\section{The Cyber Alliance Game with Power Structure}
\label{sec:power}
Our base game did not include the power distribution among the members of a given alliance. To illustrate the influence of allies, we will use the previously presented weighted voting system and the Banzhaf Power Index. Alliances are considered static, i.e., players in our model cannot change their weight value. In the present model, we differentiate between two types of alliances for visualization purposes: Dictatorial and Democratic. This allows us to examine the behavior of players with similar and significantly different alliance backgrounds. The dictatorial alliance is composed of a single Dictator and multiple Dummies such as [$10:11,2,2,2$]. The number and weight of Dummy members in the dictatorial alliance do not significantly change the voting mechanism, as the Dummy players always make decisions in line with the Dictator's interests. In the Democratic alliance, chosen to be [$20:5,5,5,5$], each player has Veto power. Naturally, an alliance might have many members and unequal weights; nevertheless, even with such changes, the behavioral characteristics typical of Veto players remain consistent with the presented findings of this study. 

\subsection{Parameters}

The meaning of parameters $p$, $q$, and $R_i$ remain consistent with their definitions in the base game. We introduce a new parameter $B_i$ to measure Player $i$'s influence within the alliance. The value of $B_i$ ranges from $0.1$ to $1.5$, and it appears as a multiplier alongside the $R_i$ parameter in the payoff function. This captures the impact of how much the Player is capable of influencing the decision regarding the use of the vulnerability after sharing it with the alliance. $B_i$ is calculated using the following formula: $B_i= 1.4 \cdot P_i + 0.1$, where $P_i$ is the Banzhaf Power Index. The addition of the $0.1$ value is necessary due to the Dummy Players, as a Dummy Player has a $0$ power index, thus it would lose any sharing reward which is not realistic. In the case of a Dictator, it results in a maximum reward increase of $+50\%$ when sharing with the alliance. For Veto players, it implies a moderate decrease because, in a Democratic alliance, all players must decide the fate of the vulnerability in the same way; accounting for the possibility that the alliance may make a decision unfavorable to the Veto player. If a Dummy player shares the vulnerability, they have very little influence over their subsequent fate and must execute the Dictator's decision, which results in losing $90\%$ of the reward.

\subsection{Equilibrium analysis}

\begin{figure}[tb]
	\centering
	\includegraphics[width=40mm, keepaspectratio]{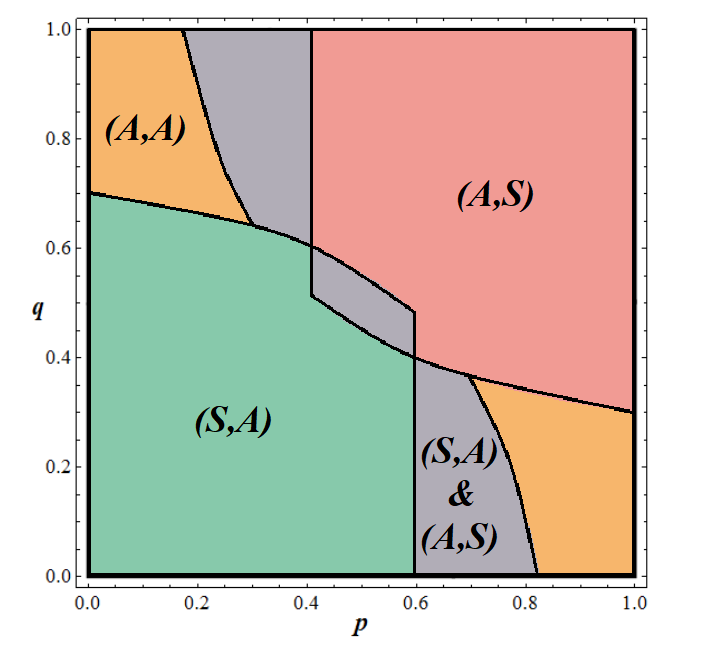}
	\includegraphics[width=40mm, keepaspectratio]{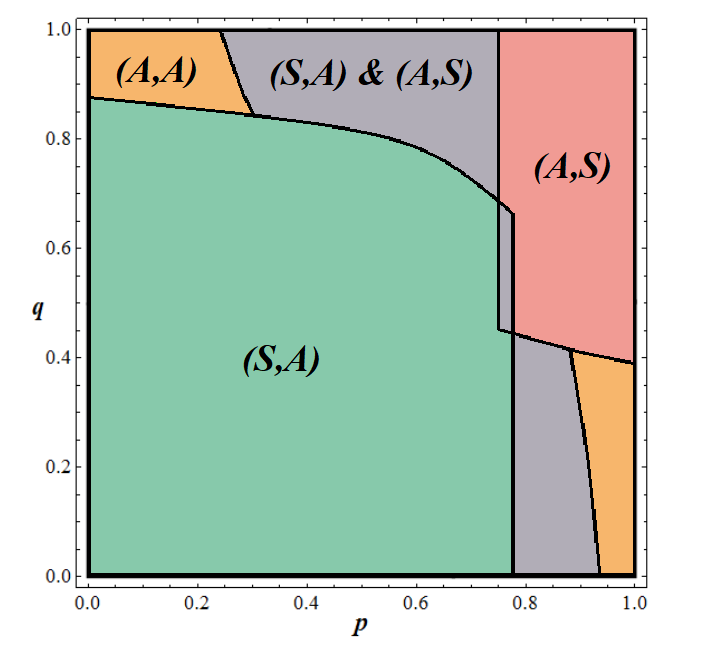}
    \includegraphics[width=40mm, keepaspectratio]{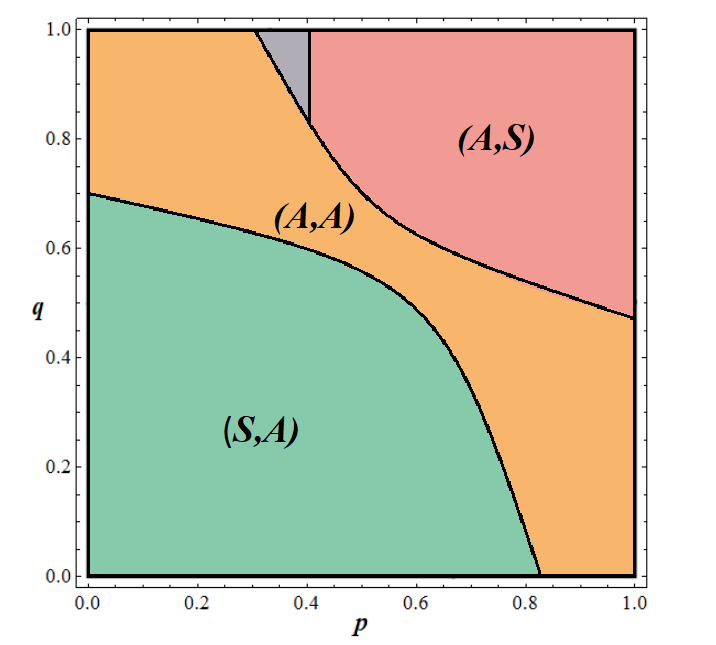}\\
\vspace{-3mm}	
 \caption{Left: Veto vs. Veto, $R_1$=$R_2$=0.9; Mid: Dict. vs. Veto, $R_1$=$R_2$=0.5; Right: Veto vs. Dummy, $R_1$=0.9, $R_2$=0.5.}
	\label{fig:power_dem}
 \vspace{-5mm}
\end{figure}

\noindent \textbf{Veto vs. Veto (Democratic vs. Democratic). }
If two Veto players from different democratic alliances face each other,  one player will almost always find sharing with the alliance to be profitable (left plot in Fig.~\ref{fig:power_dem}). Therefore, the (S,A) (green) and (A,S) (red) strategy profiles dominate when alliances reward their members appropriately. In our setting, each member's power index is $P_i=0.25$; however, real-world alliances can have various compositions. In general, the higher a player's power index, the lower the reward needs to be provided by the alliance for sharing.

\noindent \textbf{Dictator vs. Veto (Dictatorial vs. Democratic). }
In a standoff between a Dictator and a Veto player, the Veto player frequently opts for Attack, while the Dictator almost always chooses to share with the alliance (green area, middle plot in Fig.~\ref{fig:power_dem})). By increasing the reward for the Veto Player, the democratic alliance can make it less likely that the player will attack.

\noindent \textbf{Veto vs. Dummy (Democratic vs. Dictatorial). }
In the case of a Veto and Dummy player, it is not surprising that the Dummy prioritizes the Attack strategy. By increasing the reward for the Veto player, we can avoid the occurrence of the (A,A) strategy profile in cases of nearly identical technological development and aggressiveness. Naturally, increasing the reward for the Dummy player does not have an impact: it only chooses Sharing if both its technical sophistication and aggression are much lower compared to the Veto player (red area, right plot in Fig.~\ref{fig:power_dem} ).

\noindent \textbf{Dummy vs. Dummy (Dictatorial vs. Dictatorial).}
If two Dummy players from different dictatorial alliances face each other, the alliance's reward practically does not matter, regardless of how large it may be. In a dictatorial alliance, once a Dummy has shared its vulnerability, it can no longer have a say in how it will be handled going forward. Therefore, Dummies with balanced technological development and aggressiveness will find the all-attack (A,A) strategy profile optimal.

\begin{figure}[tb]
	\centering
	\includegraphics[width=55mm, keepaspectratio]{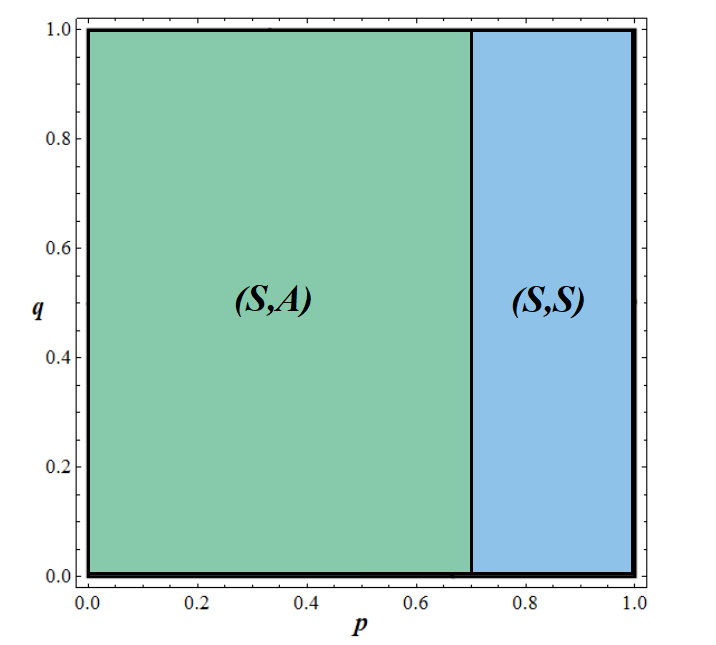}
	\includegraphics[width=55mm, keepaspectratio]{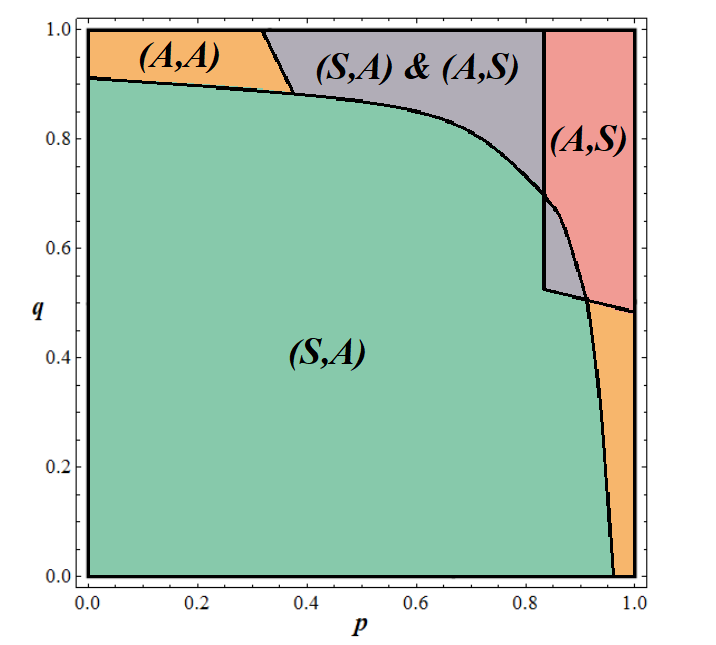}\\
    \vspace{-3mm}
	\caption{Left: Dict. vs. Dict., $R_1$=0.9, $R_2$=0.2; Right: Dict. vs. Dummy, $R_1$=0.55, $R_2$=0.4}
	\label{fig:power_dict}
    \vspace{-5mm}
\end{figure}

\noindent \textbf{Dictator vs. Dictator (Dictatorial vs. Dictatorial). }
When two leaders of dictatorial alliances face each other, even with moderate but symmetric rewards, they almost always choose to share their discovered vulnerability with their alliance. This is because they alone decide the fate of vulnerability, and the members of the alliance are obliged to accept the directive. The (S,A) profile is optimal if $R_2$ is low, and player 2 is not lagging much behind regarding technological development (green area, left plot in Fig.~\ref{fig:power_dict}). Dictators only choose the (A,A) strategy profile against each other in the case of insufficient sharing incentives from their alliance.

\noindent \textbf{Dictator vs. Dummy (Dictatorial vs. Dictatorial). }
In situations where a Dictator and a Dummy from a different alliance are clashing, it is not surprising that the Dummy will typically choose to initiate the attack, while the Dictator, even in the presence of a moderate reward, will share the vulnerability with its alliance (green area, right plot in Fig.~\ref{fig:power_dict}). Only in extreme cases, when the Dictator possesses overwhelming technological superiority and high aggressiveness, will the Dummy choose to share the vulnerability with its alliance.

\noindent \textbf{Summary. }
Their influence within their alliance greatly influences a player's behavior in specific decision-making situations. Knowing a player's power position allows for more accurate prediction of their behavior in a cyber-conflict.

Dictators wield total power within their own alliance. Discovered vulnerabilities are shared with the alliance at a very high rate, and attacks are chosen only in rare instances, as the alliance is leveraged to achieve the Dictator's objectives. Consequently, most other players find aggression to be the most appropriate response when facing a Dictator. An exception to this is when the Dictator possesses overwhelming technical and/or aggressiveness superiority. Therefore, if a Dictator wishes to avoid continuous attacks, they must demonstrate their superiority and/or aggressiveness at significant real-world financial expenditures. This is in line with the behavior of dominant leaders in both human society and the animal world.

Veto Players are driven by sharing rewards from their Democratic alliance. In case of low rewards, Veto Players are more likely to attack. Between Veto players, the (S,A) or (A,S) strategy profiles often represent an NE, reducing the likelihood of conflict escalation. Veto players with lower power might have to be compensated excessively by their alliance in order to choose not to attack.

Dummy players have no influence on alliance decisions; they essentially execute the decisions of the Dictator within the alliance. Consequently, in the vast majority of cases, Dummy players choose to attack. In fact, a Dummy may be better off outside the alliance (regarding cyber-conflicts) if they are technologically advanced (an unlikely situation).

\section{Incorporating alliance policy}
\label{sec:policy}
Alliances, in addition to the rules and guidelines outlined in their founding documents, are typically governed by numerous internal regulations and agreements. When an alliance admits a new member, the new member must adopt the alliance's guidelines and values. If this does not occur, the behavior of the alliance member reflects on the entire alliance, meaning that, e.g., launching an individual cyberattack could potentially affect all members negatively. In order to avoid such inefficiency, alliances strive to nudge members towards closer cooperation or deter them from violating the rules. In the second extension of our model, we introduce the alliance's policy, which is another tool at the alliance's disposal to influence the decisions of its members.

\subsection{Alliance posture}
In modern times, most alliances have had a defensive orientation. Their objective is not to initiate joint offensives but rather to deter enemies by demonstrating strength~\cite{nato_deterrent}. Nevertheless, we must not disregard alliances with an offensive orientation, which may arise along the lines of common interests in the event of a potentially escalated conflict or an already ongoing cyberwar.

The \emph{defensive alliance} aims to deter its members from deviating from the alliance's guidelines, in this case, deterring from exploiting a vulnerability for individual gain rather than sharing it with the alliance. Therefore, the alliance applies a penalty in the event of an attempted attack, thus reducing the payoff of the deviating member state. On the other hand, the \emph{offensive alliance} evaluates the actions of its members, as this type of alliance typically forms during an ongoing or emerging conflict. Therefore, the offensive-oriented alliance also rewards successful attacks.

\begin{figure}[tb]
	\centering
	\includegraphics[width=0.63\textwidth]{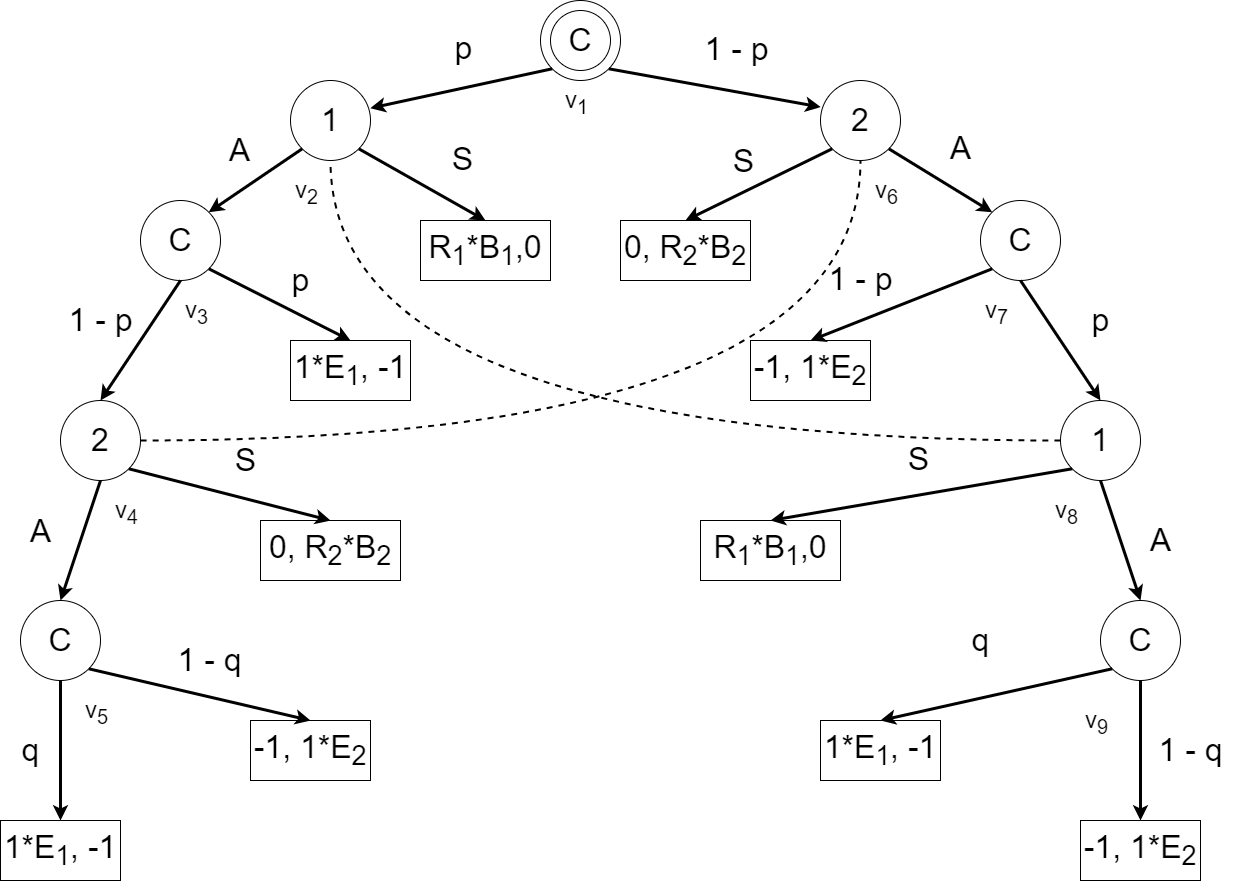}
	\caption{Cyber Alliance Game with both power and policy structure in extensive form}
	\label{fig:game2}
\end{figure}

\subsection{Parameters}
The meaning of parameters $p$, $q$, $R_i$, and $B_i$ remain consistent with their definitions established in the Cyber Alliance Game with Power Structure. We introduce a new parameter $E_i$ that conveys the alliance's punishment/reward on the payout generated by an individual cyberattack. For an offensive alliance, the values of $E_i$ can range from $1.1$ to $1.6$ (reward), while for a defensive alliance, this value ranges from $0.9$ to $0.4$ (penalty), i.e., a reward or penalty of $\pm 10\%$ to $\pm 60\%$. Note that these value sets were determined only for demonstration purposes. The resulting extended game tree is shown in Fig.~\ref{fig:game2}.

\subsection{Equilibrium analysis}

\noindent \textbf{Dictator vs. Dictator. }
\begin{figure}[!tb]
	\centering
        \includegraphics[width=55mm, keepaspectratio]{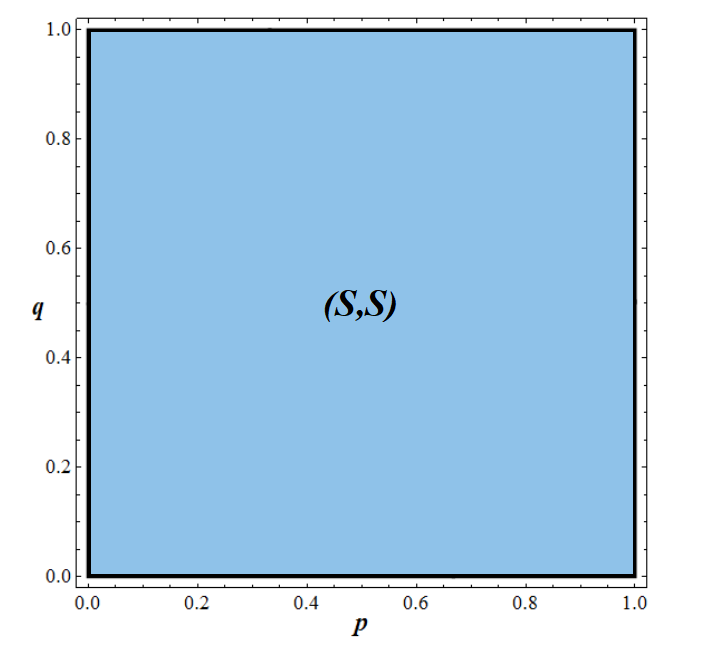}
	\includegraphics[width=55mm, keepaspectratio]{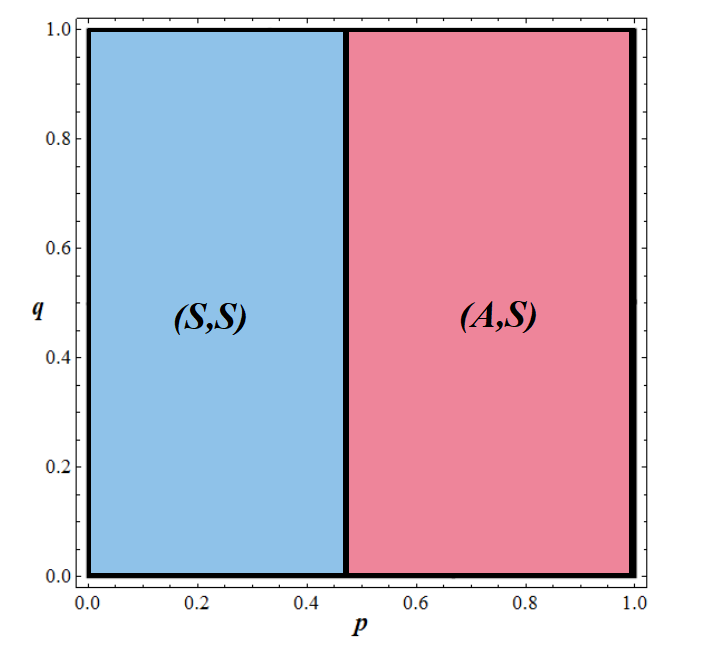}\\
 \vspace{-3mm}
	\caption{Left: Defensive Dictator vs. Defensive Dictator ; Right: Defensive Dictator vs. Offensive Dictator}
	\label{fig:ddictddictkozepeskeszandddictodictkozepeskesz}
  \vspace{-5mm}
\end{figure}
In the case of Dictators, we could see in the first extension that sharing with the alliance often represents the most profitable decision for the players; they alone can decide as Dictators about the further fate of the vulnerability while also pocketing some sharing rewards. Therefore, the punishment for an attack in the case of a Dictator only encourages more sharing with the alliance (see the left plot in Fig.~\ref{fig:ddictddictkozepeskeszandddictodictkozepeskesz}). In contrast, influencing the Dictator's decision is possible with the reward for the attack. If the attack is sufficiently rewarded and the sharing reward of the alliance is low, then the Dictator Player is more likely to attack, especially when being technologically more advanced than its enemy (red area, right plot in Fig.~\ref{fig:ddictddictkozepeskeszandddictodictkozepeskesz}).

\noindent \textbf{Veto vs. Dummy. }
\begin{figure}[tb]
	\centering
	\includegraphics[width=55mm, keepaspectratio]{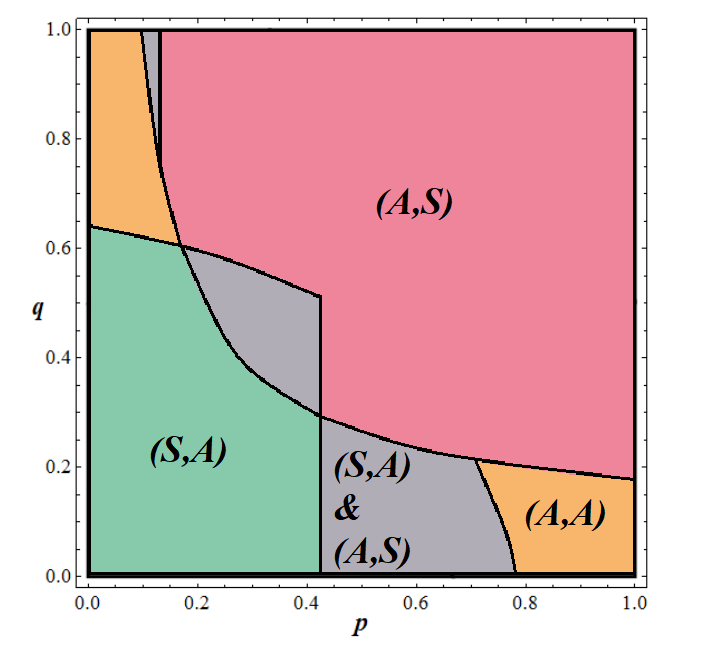}
	\includegraphics[width=55mm, keepaspectratio]{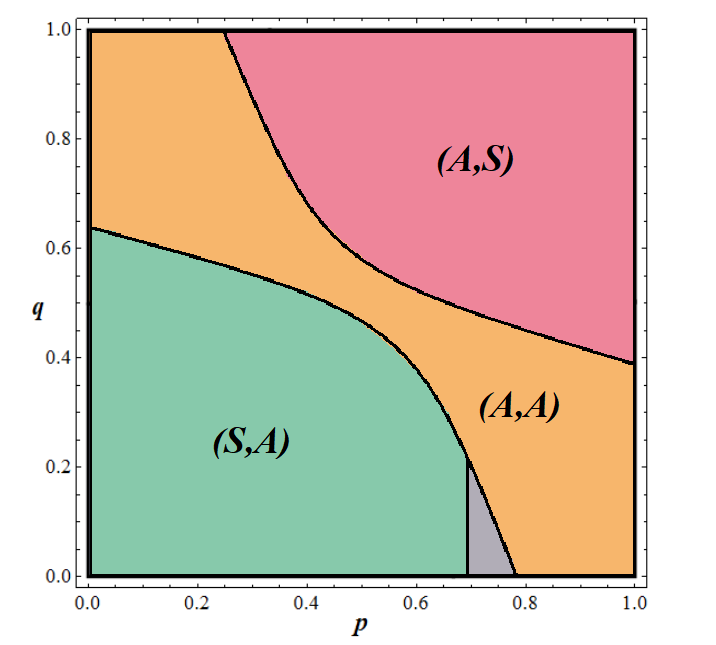}\\
    \vspace{-3mm}
	\caption{Left: Defensive Dummy vs. Defensive Veto; Right: Defensive Dummy vs. Offensive Veto}
	\label{fig:ddummydvetomagaskeszandddummyovetomagaskesz}
    \vspace{-5mm}
    \end{figure}
The actions of Veto players can be influenced to the greatest extent by the reward or punishment for an attack. Through the reward for sharing and the reward/punishment for an attack, the alliance is almost fully capable of controlling the decisions of Veto members in certain situations. With a low reward for sharing and a high reward for attacking, Veto Players can be encouraged to attack, while punishing the attack and rewarding high sharing can persuade them to cooperate with the alliance (see Fig.~\ref{fig:ddummydvetomagaskeszandddummyovetomagaskesz}). Note that we chose to show Dummy vs. Veto to isolate the effects of alliance policy on the Veto player.

\noindent \textbf{Dummy vs. Dummy. }
\begin{figure}[tb]
	\centering
	\includegraphics[width=55mm, keepaspectratio]{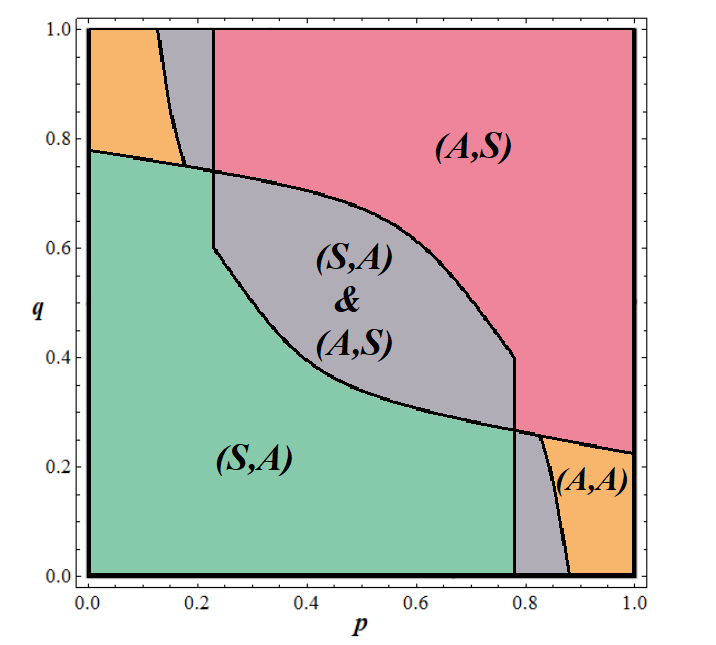}
	\includegraphics[width=55mm, keepaspectratio]{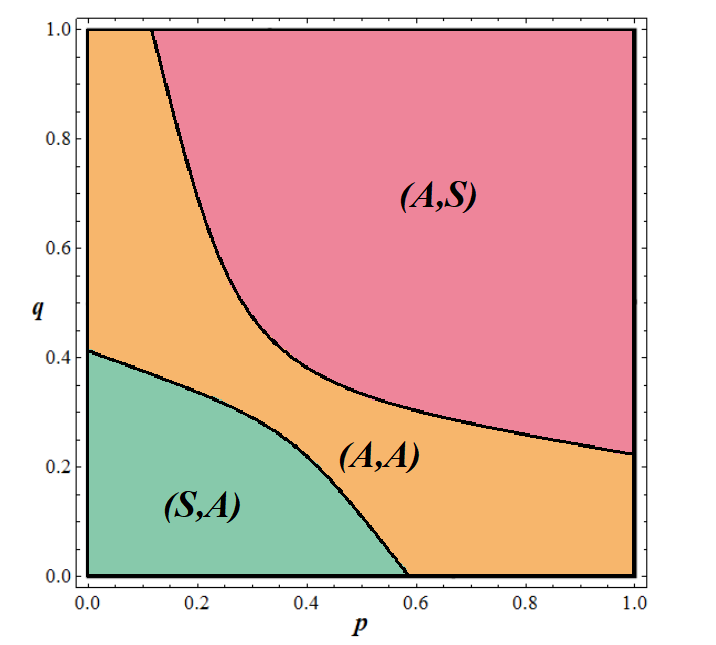}\\
  \vspace{-3mm}
	\caption{Left: Defensive Dummy vs. Defensive Dummy; Right:  Offensive Dummy vs. Defensive Dummy}
	\label{fig:ddummyddummymagaskeszandodummyddummymagaskesz}
  \vspace{-5mm}
\end{figure}
As we have seen earlier, Dummy Players most often choose to attack because sharing their discovered vulnerability with the alliance brings them very little benefit. They have no influence over how the vulnerability is utilized, and as members of the alliance, they must accept the decision. Therefore, Dummies can be best persuaded to cooperate with the alliance via a stringent penalty for attacking. However, an all-sharing equilibrium emerges very rarely (see Fig.~\ref{fig:ddummyddummymagaskeszandodummyddummymagaskesz}), as at least one of the Dummies will achieve a higher payoff attacking (with the exception of very high attack penalties).

\noindent \textbf{Summary. }
Changing the payout for attacks is a useful tool for alliances to influence the decisions of their members in certain situations. Dictators in an offensive-oriented alliance are more likely to attack than in a defensive alliance. Veto players can be greatly controlled by the alliance's ability to change the payout for attacks. Dummies can be encouraged to cooperate based on punishments for attacks. 

By influencing the payout for attacks, alliance members are more exposed to the influence of their alliance. In a dictatorial alliance, the Dictator can shape the decisions of Dummy players according to their preferences. If the alliance provides a low reward for sharing but supports attacking, Dummy players will invariably choose the attack. On the contrary, with high attack penalties and high alliance rewards, they will choose to share. Thus, Dummy players are even more dependent on the will of the Dictator in such situations.

\section{Conclusion}
\label{sec:conclusion}
The Cyber Alliance Game demonstrates that alliances can incentivize their members to cooperate with the alliance through rewards, as cooperation without such incentives would only occur in extreme cases, given that advantages achievable through attacks are only possible through payouts. However, members may have varying degrees of influence within the alliance. Introducing the concept of power to The Cyber Alliance Game with Power Structure, derived from the Banzhaf Power Index, allows us to model the differences among members. Dictator, Veto, and Dummy players have unique roles and fates within the life of an alliance, which can be exploited by either the opposing parties or the allies themselves. Ultimately, by considering the possibility of changing the payouts resulting from attacks, alliances can exert even greater control over their members. Defensive and offensive alliances could change the default behavior of their member states to the alliance's own benefit by punishing or rewarding individual attacks. In real life, knowing the opponent well, understanding how their alliance operates, and their position within the alliance can help predict the decisions of a given nation-state.

\subsection{Democratic defensive alliances}
In a Democratic alliance, the majority of members respond well to alliance rewards and punishments, making defensive behavior sustainable for almost every member in practice. The alliance democratically decides on the exploitation of vulnerabilities, making vulnerability sharing favorable for members. Such an alliance is most useful during peacetime, as the balance can be maintained by adjusting reward parameters. If members engage in too many individual actions, the trend can be reversed by increasing penalties for attacks and/or increasing rewards for sharing. On the other hand, if members share vulnerabilities excessively, meaning they do not carry out individual actions and overly direct their local problems toward the alliance globally, there is an opportunity to reduce the punishment for attacks.

During peacetime, maintaining and joining the Democratic Defensive alliance is the optimal decision for a Veto player. The power values of Veto players may vary a little but not much; consequently, the alliance's response to changes in rewards and punishments is almost uniform, rendering the alliance predictable. The behavior of alliance members formed by Veto players can be shifted towards a higher proportion of sharing and a higher proportion of attacks, although this incurs relatively high costs for the alliance. Therefore, Veto players in Democratic alliances respond well to changes in parameters, but the alliance may only divert them from their individually preferred behavior with very high rewards or punishments; this strengthens the sovereignty of member states.

\subsection{Dictatorial offensive alliances}
The formation of a Dictatorial Offensive alliance is usually advisable in cases of pre-existing conflicts. The alliance founder is typically the Dictator, who gathers weaker members into the alliance. Since members do not receive high rewards through sharing, they will constantly choose to attack, engaging in individual conflicts. In contrast, the Dictator almost always shares vulnerabilities with the alliance, forcing members to act based on the Dictator's decision. If a Dictator discovers a very large number of vulnerabilities or a critical weakness, it is worth sharing them with the alliance, as this can prompt all members to engage in a large-scale coordinated cyberattack.

During peacetime, it is recommended to incentivize Dummy players to better align with the alliance's interests. However, this incurs significant costs. To address this, the Dictator must either increase the penalty stakes or, during peace, implement some form of democratization within the alliance, a subject not explored in this paper. Additionally, the Dictator must be vigilant regarding technological advancements or high levels of aggressiveness, as most adversaries would likely target the Dictator, given the knowledge that they almost always refrain from an individual attack. Therefore, the role of the Dictator is highly effective in an established conflict from the Dictator's perspective but can be prohibitively costly during peacetime.


\subsection{Limitations and future work}
We made several simplifications in order to i) provide tractable models and ii) be able to compare to the baseline Cyber Hawk game~\cite{moore2010would}. 
Attacks are always assumed successful, and there is no potential for retaliation against the attacker. The game is myopic, and we restrict the number of players to two. We did not delve into how the alliance generates player rewards, and we did not model the alliance's decision after a vulnerability was shared. We restricted the alliances to two fixed types, Democratic and Dictatorial, although a mixed form may occur with an infinite number of combinations for member count and power index. Furthermore, when it comes to penalties in defensive alliances, we did not consider potential negative payouts or the exclusion of a player from the alliance. In future work, we plan to relax our assumptions and characterize long-sighted equilibria with repeated interactions, the potential for retaliation, heterogeneous alliances, and dynamic alliance posture.

\noindent \textbf{Acknowledgements. }This work has been funded by
Project no. 138903, implemented with the support provided by the Ministry of Innovation and Technology from the National Research, Development, and Innovation Fund, financed under the FK\_21 funding scheme.

\appendix
\section*{Appendix}
\section{Real-world cyber-warfare}
\label{sec:real}
Cyberattacks can be broadly categorized into two types: they could target military or civilian objects. Military targets entail the leadership and command of armed forces, as well as weapon guidance systems. Civilian targets encompass critical infrastructure, administrative systems, banking systems, the private sector, and high-ranking political or economic leaders. In the current Ukrainian conflict, we are witnessing an unprecedented, prolonged, intensive cyberwar between two nation-states; even involving civilian volunteers in the operations. In the following, we will examine the largest cyberattacks against civilian targets during the Russo-Ukrainian war.

\noindent \textbf{Ukraine power grid. }In December 2015 and again in December 2016, Ukraine's electrical grid experienced severe cyberattacks \cite{case2016analysis}. In both instances, the Supervisory Control and Data Acquisition (SCADA) system of the electricity transmission company was targeted. While the infrastructure of the electrical network was not physically harmed during the attack, approximately $200,000$ people were left without service for several hours. Due to their long lifespan, SCADA systems often carry unpatched vulnerabilities and become targets of attacks~\cite{alanazi2023scada}. 
 
\noindent \textbf{Kyivstar. }The largest cyberattack in the telecommunications sector was suffered by Kyivstar~\cite{santora2023huge}. The company, which had 24 million users, saw its phone and internet services and website become completely inaccessible due to the attacks. In some areas, even air raid alert systems became inoperative. Additionally, the attackers caused significant damage to the company's systems, particularly concerning cloud-based virtual environments.

\noindent \textbf{VTB. }Russia's second-largest bank, VTB, was subjected to the largest DDoS attack in history\footnote{\url{https://www.reuters.com/business/finance/russian-state-owned-bank-vtb-hit-by-largest-ddos-attack-its-history-2022-12-06/}}, during which most of its services, including the mobile app and website, became inaccessible. The company issued a brief statement informing its customers about the attack originating from ``abroad''.

\noindent \textbf{IPL Consulting. }Russia's largest and most advanced company specializing in the implementation of industrial IT systems was hit by a cyberattack~\footnote{\url{https://kyivindependent.com/military-intelligence-claims-cyberattack-on-russian-defense-ministry-gave-access-to-classified-documents/}}. The attackers infiltrated the company's IT systems, destroyed vast amounts of data, and rendered numerous servers and services inoperative. The company's clients include major companies in the automotive, aerospace, and defense industries.

\section{(Cyber-)alliances}
\label{sec:alliances}
To complicate matters, most countries are part of some form of military or economic alliance. These alliances usually have strict admission criteria, internal rules, and a common purpose. The primary objectives of these alliances used to be defensive. Once a nation-state becomes a member of an alliance, it represents both its own interests and those of the alliance through its actions and behavior. If a member is under attack, the other members typically come to their aid according to internal regulations. Earlier, during collaborations, members aligned their military and economic objectives, and nowadays, a common cyber defense strategy has also been implemented in most alliances, blurring the lines between military and other alliances. The alliances presented below effectively demonstrate the diverse objectives, power structures, and internal regulations they can operate with.

\noindent \textbf{European Union. }The European Parliament, the Council of the European Union, and the European Commission form the legislative branch of the alliance. The EU has been advocating for harsh sanctions against cybercrime, the detection of abuse of non-paper-based payment instruments, the introduction of quantum encryption, as well as joint research and knowledge sharing \cite{christou2018challenges}. A recent barrage of new cybersecurity-related regulations, including the Network and Information Security (NIS2), the Digital Operational Resilience Act (DORA), the Cyber Resilience Act (CRA), and others \cite{vandezande2024cybersecurity}, compulsory to adopt in member states, ensure that cybersecurity (including building defensive capacity against cyberattacks, swiftly patching vulnerabilities, and reporting cyber-incidents to the national and EU authorities with a short deadline) is a first-order priority.

\noindent \textbf{BRICS+. }In 2006, Brazil, Russia, India, and China created the ``BRIC'' alliance. The group was designed to bring together the world's most important developing countries and challenge the political and economic power of the wealthier nations of North America and Western Europe. Since its founding, 5 additional countries have joined the alliance, now referred to as BRICS+, with several other prospective members and interested countries. The economic alliance encompasses more than $44\%$ of the world's population and over $28\%$ of the world's economic power. The member countries, within the framework of the CyberBRICS project~\cite{belli2021cyberbrics}, aim to create a uniformly structured and secure cyberspace. CyberBRICS is an international research project with three main areas: protection of personal data, secure digital transition, and regulation of artificial intelligence~\cite{cyberbrics}.

\noindent \textbf{North Atlantic Treaty Organization (NATO). }
As the military alliance with the most member countries, NATO has prioritized the importance of information systems defense~\cite{hunker2010cyber}. In 2012, NATO decided to centralize the protection of all communication networks within the alliance. Following a decision in 2016, NATO began to intensify its cooperation with the private sector, leading to the inclusion of ``NATO-compatible'' products among the top-tier offerings of most major manufacturers. The highest political decision-making body is the North Atlantic Council, where each member country is represented. Decisions affecting the operation of the alliance, such as admitting a new member, require the consent of every member country. However, the level of technological advancement, the number of experts, the amount of resources allocated, etc., vary significantly across nations, resulting in widely differing power positions within the alliance. The alliance employs a policy of deterrence through cyber exercises and enhancing their capabilities~\cite{elamiryan2018comparing}.

\noindent \textbf{Collective Security Treaty Organization (CSTO). }The CSTO is a military alliance consisting of five former Soviet successor states alongside Russia. The organization's operations, regulations, and guidelines have received less publicity compared to NATO~\cite{elamiryan2018comparing}. Article 4 of the CSTO charter adopted in 2002 states that if one of the Member States undergoes aggression (armed attack menacing to safety, stability, territorial integrity, and sovereignty), it will be considered by the Member States as aggression to all the Member States of this Treaty. Neither the Article nor other agreements address how the alliance handles cyberattacks. Generally speaking, the organization pays significant attention to the development of cyber capabilities and the use of ``hybrid technologies''. Additionally, there is a significant willingness for cooperation among the countries to combat domestic cybercrime and impose stricter cyber hygiene.

\noindent \textbf{Five Eyes. }The alliance consisting of 5 Anglo-Saxon countries is one of the most successful intelligence cooperations since World War II~\cite{wells2020between}. The main area of cooperation among the countries is Signals Intelligence (SIGINT), within which they share information collected from adversarial countries. The key to successful collaboration is global reach, high technological readiness, similar legal systems, and, last but not least, a common language and culture. In most cases, the members follow and adhere to decisions made by their allies in a predictably disciplined manner. However, certain steps can cause tension and even public debate among the member countries. Recently, a divisive issue was the restriction of products from Chinese companies Huawei and ZTE~\cite{shoebridge2018chinese}.

\bibliographystyle{splncs04}
\bibliography{bib/mybib}

\end{document}